\documentclass[preprint,12pt,authoryear]{elsarticle}

\ifx\pdfoutput\undefined
\usepackage[dvips]{graphicx}
\else
\usepackage{graphicx}
\fi
\usepackage{epstopdf}
\usepackage{floatrow}

\newfloatcommand{capbtabbox}{table}[][\FBwidth]

\usepackage{multirow} 
\usepackage{amsmath}
\usepackage{amssymb}
\usepackage{amsthm}
\usepackage{mathtools}
\usepackage{amsfonts}
\usepackage{natbib}
\usepackage{caption}
\usepackage{multirow}
\usepackage{soul}
\usepackage[T1]{fontenc}
\usepackage{blindtext}
\usepackage{epsfig}
\usepackage{bm}
\usepackage{booktabs}
\usepackage{float}
\usepackage{algorithmic}

\usepackage{float}
\usepackage{subcaption}
\usepackage[ruled,vlined]{algorithm2e}
\floatstyle{plaintop}
\restylefloat{table}
\usepackage{graphicx}
\usepackage[utf8]{inputenc}
\usepackage[english]{babel}
\usepackage[normalem]{ulem}

\usepackage[font=small,labelfont=bf]{caption}\usepackage[pdftex,dvipsnames,usenames]{color}
\usepackage{sectsty}
\allsectionsfont{\usefont{OT1}{phv}{bc}{n}\selectfont} 

\DeclareMathOperator*{\argmin}{arg\,min}

\long\def\symbolfootnote[#1]#2{\begingroup%
	\def\thefootnote{\fnsymbol{footnote}}\footnote[#1]{#2}\endgroup}

\begin{document}

\begin{frontmatter}

\title{Applying Regression Conformal Prediction with Nearest Neighbors to time series data}
\author[1]{Samya Tajmouati
\corref{cor1}}
\ead{samya.tajmouati@gmail.com}

\author[1]{ Bouazza El Wahbi\corref{cor1}}

\author[2]{Mohammed Dakkoun\corref{cor1}}

\address[1]{Department of Mathematics, Ibn Tofail University, Faculty of Sciences, K\'enitra, Morocco}

\address[2]{The research team of Modeling and Information Theory, Abdelmalek Essa\^{a}di University, T\'etouan, Morocco}

\begin{abstract}

In this paper, we apply conformal prediction to time series data. Conformal prediction is a method that produces predictive regions given a confidence level. The regions outputs are always valid under the exchangeability assumption. However, this assumption does not hold for the time series data because there is a link among past, current, and future observations. Consequently, the challenge of applying conformal predictors to the problem of time series data lies in the fact that observations of a time series are dependent and therefore do not meet the exchangeability assumption. This paper aims to present a way of constructing reliable prediction intervals by using conformal predictors in the context of time series. We use the nearest neighbors method based on the fast parameters tuning technique in the weighted nearest neighbors (FPTO-WNN) approach as the underlying algorithm.  Data analysis demonstrates the effectiveness of the proposed approach.

\end{abstract}

\end{frontmatter}
\makeatletter
\def\ps@pprintTitle{%
  \let\@oddhead\@empty
  \let\@evenhead\@empty
  \def\@oddfoot{\reset@font\hfil\thepage\hfil}
  \let\@evenfoot\@oddfoot
}
\makeatother

\section{Introduction}
Many traditional machine-learning algorithms focus on providing point forecasts. However, in many cases, it is necessary to output the regions in which the unknown values should fall. As a result, prediction intervals should be examined \citep{vovk2005conformal,balasubramanian2014conformal,kowalczewski2019normalized}. To provide prediction intervals, some machine learning methods do exist where conformal prediction is considered one of the most popular methods that provides valid regional predictions \citep{vovk2005conformal, papadopoulos2011regression, balasubramanian2014conformal}. Besides, unlike other machine learning methods such as the Bayesian approach, the application of conformal prediction requires only the exchangeability assumption.
We say that $(x_1, x_2, \cdots)$ is an infinitely exchangeable sequence of
random variables if, for any $n$, the joint probability $p(x_1, x_2, . . . , x_n)$ is invariant to permutation of the indices. That is, for any permutation $\pi$,
\[
p(x_1,x_2,\cdots,x_n) = p (x_{\pi_1}, x_{\pi_2}, \cdots, x_{\pi_n})
\]

The conformal prediction technique has been successfully applied to the regression and classification tasks. It has, for example, been applied to  regression conformal prediction with nearest neighbours \citep{papadopoulos2011regression}; conformal regression forests \citep{bostrom2017accelerating}; binary classification of imblanced datasets \citep{norinder2017binary}; and an electronic nosebased assistive diagonostic protype for lung cancer detection \citep{zhan2020electronic}. However, few works address the application of conformal prediction to time-series data. For instance, \cite{kath2020conformal} employ the conformal prediction technique for short-term electricity price forecasting. They show that conformal prediction yields reliable prediction intervals in short-term power markets after combining it with various underlying forecasting models. \cite{kowalczewski2019normalized} places some assumptions, which convert time series data into regression problem in which conformal prediction is applicable. \cite{dashevskiy2011time} consider the problem 
3
 of applying conformal predictors to time series prediction in general and the network traffic demand prediction problem in particular. They show that in the case when the time series data do not meet the requirement of exchangeability, conformal predictors provide reliable prediction intervals, which indicates empirical validity. Moreover, they test different point forecasts algorithms in order to determine the points that are empirical efficient. \cite{balasubramanian2014conformal} consider an application of conformal prediction to the network traffic demand time series and show the empirical validity of the conformal prediction when the time series are not exchangeable. 

Following the work of \cite{dashevskiy2011time} and \cite{balasubramanian2014conformal}, we extend the application of conformal predictors to time series data when the exchangeability assumption is not met. Our approach differs from the existing one by the fact that it builds prediction regions in the multidimensional case and employs as the underlying algorithm the WNN method based on the FPTO method as described in \citep{tajmouati2021}.Also, we extend our approach such that it provides efficient point forecasts, which lead to efficient prediction intervals \citep{tajmouati2021}. Our method shows promising results. 

The paper is organized as follows. Section 2 introduces the conformal prediction method. Sections 3 and 4 describe the general idea behind the FPTO-WNN method and the application of the conformal prediction to time series with the FPTO-WNN approach, respectively.  A simulation study and an example with real data are provided in Section 5. Section 6 contains a summary.


\section{Conformal prediction}
\label{sec.2}
Conformal prediction approach uses past experience to determine precise levels of confidence in new predictions. It is a method introduced first by \cite{vovk1998} and studied by \cite{vovk2005conformal} and \cite{shafer2008tutorial}. Its main objective is to produce valid prediction intervals under the exchangeability assumption of the data. The original version, which was implemented in the transductive manner, is defined as follows. Given an exchangeable sequence $z_1,z_2,\cdots,z_l$ where $z_i=(x_i,y_i )$ is the $i^{th}$ pair such that $x_i\in \mathbb{R}^d$ and $y_i\in \mathbb{R}$ are the object and the label, respectively.
Given a new unlabeled example $x_{l+1}$, the task of the conformal prediction is to output an interval that contains label $y_{l+1} $ of $x_{l+1}$. 
To do this, one has to define some non conformity measure: $A_{l+1}: Z^{l}\times Z \to \mathbb{R}$, which attributes a numerical score $\alpha_i = A_{l+1} \left(\left[ z_1,z_2,\cdots,z_{i-1},z_{i+1},\cdots,z_{l+1}\right],z_i \right) $ to each example $z_i$ measuring the degree of disagreement between its label $ y_i $ and the predicted label $\hat{y _i}=D_{\{z_1,z_2,\cdots,z_{i-1},z_{i+1}, \cdots,z_{l+1} \}} \left(x_i\right)$, where $ D_{\{z_1,z_2,\cdots,z_{i-1},z_{i+1}, \cdots,z_{l+1} \}}$ is the prediction rule created by the underlying algorithm used to predict $y_i$ based on all the examples except  $z_i$. The degree of disagreement measures the deviation between the predicted value and the observed value. That is, if the predicted value is close to the observed one, the disagreement is considered weak. In the conformal prediction, the underlying algorithm means the algorithm like Neural Network, SVM, KNN, and regression, to name a few that provides point forecasts, which are later used to generate a prediction interval.
Usually, one uses the function $|y-\hat{y}|$ as a non-conformity score. The non-conformity score $ \alpha_{l+1} $ does not reflect on its own any information and need to be compared with all other non-conformity measures. This comparison can be performed through the p-value function calculated for the new example $z_{l+1}$ as:
\begin{equation}
p\left(\tilde{y}\right)= \frac{\#{\{i=1,\cdots,l+1 : \alpha_i \geq \alpha_{l+1} \} }}{l+1}
\label{eq.1}
\end{equation}
According to Eq. (\ref{eq.1}), the higher the value of $ \alpha_{l+1} $,  the less probable it is that $ y_{l+1} $ takes $ \tilde{y} $. Finally, given a significance level $ \delta $, a regression conformal predictor outputs the following prediction region:
\begin{equation}
\Gamma = \{\tilde{y} : p\left(\tilde{y}\right) > \delta \} 
\label{eq.2}
 \end{equation}

To overcome the computational inefficiency in the transductive conformal prediction, an inductive conformal prediction is proposed. The general steps in the inductive approach are as follows:

\begin{itemize}
\item Split the training dataset $z_1, z_2,\cdots, z_l$ into two smaller datasets: the calibration dataset with $q < 1$  examples and the proper training dataset with m:=l-q examples.

\item Use the proper training data $z_1,z_2,\cdots,z_m$ to generate the prediction rule $D_{\{z_1,z_2,\cdots,z_m\}}$ created by the underlying algorithm. 

\item Attribute a non-conformity score to each one of the examples in the calibration set. Note that the non-conformity score $\alpha_{m+i}$ of each example $z_{m+i}$ in the calibration dataset $z_{m+1},z_{m+2},\cdots,z_{m+q}$ is calculated as the degree of disagreement between $\hat{y}_{m+i}=D_{\{z_1,z_2,\cdots,z_m \}}\left(x_{m+i}\right)$ and the real value $ y_{m+i}$.

\item Define p-value of $\tilde{y}$ of $ x_{l+g}$ as:
\[p\left(\tilde{y}\right)=\frac{\#{\{i= m+1,\cdots, m+q, l+g :\ \alpha_i \geq \alpha_{l+g} \} }}{q+1}, \]
where $\alpha_{l+g}$ is the degree of disagreement between $\hat{y}_{l+g}=D_{\{z_1,z_2,\cdots,z_m \}}\left(x_{l+g}\right)$ and $\tilde{y}$.

\item Sort the non conformity scores of the calibration examples $ \alpha_{m+1},\cdots,\alpha_{m+q} $ in descending order: $\alpha_{\left(m+1\right)},\cdots, \alpha_{\left(m+q\right)}$.

\item For a significance level $\delta$ and $\alpha_i = |y_i-\hat{y}_i|$, output the prediction region as: 

\[\Gamma = \{\tilde{y} : p\left(\tilde{y}\right) > \delta \}:= \left]\hat{y}_{l+g} - \alpha_{\left(m+s\right)} ; \hat{y}_{l+g} +\alpha_{\left(m+s\right)}\right[,\]

where $s = \lfloor \delta\left(q+1)\right) \rfloor$. 
\end{itemize}

\section{FPTO-WNN}
\label{sec.FPTO}
FPTO-WNN is an automatic method that selects the optimal values of the nearest neighbors and performs feature selection in the weighted nearest neighbors (WNN) approach for time series data. The method works as follows. Let $a=\left(a_1,a_2,\cdots,a_T\right) $ and $ E=\{E_1,\cdots,E_I \}$ be a time series and a set of $I$ training datasets of a, respectively. For $i \in \{1,\cdots,I\}, E_i=\{a_1,a_2,\cdots,a_{T-i.n}\}$ is the $i^{th}$ training dataset used to forecast the $i^{th}$ test dataset: $a_{T-i.n+1},\cdots,a_{T-i.n+n} $. The optimal values of p and k are denoted by $p^*,k^*$, respectively, and are the values that minimize
\[
\begin{split}
\text{MAPE} ^*\left(p,k\right) : \left(p^*,k^*\right) &=\argmin \text{ MAPE}^* \left(p,k\right), \text{ where}\\
\text{MAPE}^* \left(p,k\right) &= \frac{1}{I}\sum_{i=1}^{I}\frac{100}{n}\sum_{j=1}^{n}|\frac{\hat{a}_{T-i.n+j}-a_{T-i.n+j}}{a_{T-i.n+j}}|. 
\end{split}
\]
$\hat{a}_{T-i.n+j}$ is calculated according to the WNN approach. MAPE is the mean absolute percentage error. The interested reader may refer to the work of \cite{tajmouati2021} paper for further theoretical analysis. Overfitting can be identified by splitting the time series data into training and testing datasets. The training dataset is split into a new train dataset and a validation dataset. Then, the model is iteratively trained and validated on the new train and validation sets according to the FPTO-WNN approach. Finally, the performance of the obtained model on the test dataset is assessed.

\section{Conformal prediction for time series with FPTO-WNN}
\label{sec.new method}
Section \ref{sec.FPTO} presents an efficient way to provide point predictions for time series data. However, these point predictions are not associate with confidence information. In this section, we introduce a method that builds the prediction region for the point predictions for time series data. The conformal prediction is combined with the FPTO-WNN approach and works as follows. First, the time series $a = \left(a_1,a_2,\cdots,a_T\right)$ is transform to pairs : 
\[z_t = \left(\text{object, label}\right) = \left(x_t,y_t\right):= \left(\left(a_{t-n.p+1}, a_{t-n.p+2},\cdots, a_t\right),\left(a_{t+1}, \cdots, a_{t+n}\right)\right),\]
for $t = \{T-n, T-2.n,\cdots, T-n.c\}$, where p is an integer and c is the last integer that holds the following inequality $T-n.c \geq n.p$. Then, the prediction region of $y_T = \left(a_{T+1},\cdots,a_{T+n}\right)$ is calculated by defining the non-conformity measure based on the WNN approach as defined in Section \ref{sec.2}. The implemented non-conformity scores are defined as follows: \[
\alpha_{t,j} = |a_{t+j}-\hat{a}_{t+j}|  \text{ for } j=1,2,\cdots,n,
\]
where $\hat{a}_{t+j}$ is the $j^{th}$ component of $ \hat{y}_t=D_{\{z_{t-n},\cdots,z_{T-n.c} \}} \left(x_t\right)$, such that $ D_{\{z_{t-n},\cdots,z_{T-n.c} \}}$ is the prediction rule created by the WNN's approach using all the previous examples of $z_t$. For implementation, we consider the following scheme: 

\begin{itemize}
\item Calculate the non-conformity scores: $ \alpha_{T-n,j},\alpha_{T-2.n,j},\cdots,\alpha_{T-h.n,j} $, for $j=1,\cdots,n$, where h is an integer inferior or equal to c.

\item Define p-value of $\tilde{y}_{T,j}$ for $x_T$ as : 
\[ p\left(\tilde{y}_{T,j}\right)= \frac{\#{\{t=T-n, T-2.n,\cdots,T-h.n,T :\ \alpha_{t,j} \geq \alpha_{T,j} \} }}{h+1}.\]

\item For each $j= 1,\cdots,n$, sort the non conformity scores: $\alpha_{T-n,j},\alpha_{T-2.n,j},\cdots,\alpha_{T-h.n,j}$ in descending order obtaining the sequences : $ \alpha_{\left(T-n,j\right)},\alpha_{\left(T-2.n,j\right)},\cdots,\alpha_{\left(T-h.n,j\right)}$

\item For a significance level  $\delta $, output the prediction interval: 
\[ \Gamma_{T,j} = \{\tilde{y}_{T,j} : p\left(\tilde{y}_{T,j}\right) > \delta \}:= \left]\hat{a}_{T+j} - \alpha_{\left(T-s.n,j\right)} ; \hat{a}_{T+j} + \alpha_{\left(T-s.n,j\right)}\right[,\]
where $s = \lfloor \delta\left(h+1)\right) \rfloor$.

\item For a significance level $\delta$, output the prediction region: \[\Gamma_T = \prod_{j=1}^{n}\Gamma_{T,j} = \prod_{j=1}^{n}\left]\hat{a}_{T+j} - \alpha_{\left(T-s.n,j\right)}; \hat{a}_{T+j} + \alpha_{\left(T-s.n,j\right)}\right[.\]
\end{itemize}

\textbf{\textit{Checking the validity and efficiency of the proposed method:}}

\begin{algorithm}[!ht]
\caption{$\text{CheckCP}\left(a, p^*, n, k^*, I_1, I_2, \delta\right)$}
\begin{algorithmic}
\REQUIRE Time series $\left(a\right)$, Window's length $\left(p^*\right)$, Number of predictions to be made $\left(n\right)$, Number of neighbors $\left(k^*\right)$, Number of validation sets used in FPTO-WNN $\left(I_1\right)$, Number of test examples $\left(I_2\right)$, significance level $\delta$ 
\vspace{0.25cm}
\STATE $M \leftarrow$ empty matrix with n columns and $I_2$ rows
\STATE $P \leftarrow$
$\begin{pmatrix}
\alpha_{T-nI_1-nI_2,1} &\cdots  & \alpha_{T-nI_1-nI_2,n}\\
\alpha_{T-nI_1-nI_2+n,1}  &\cdots & \alpha_{T-nI_1-nI_2+n,n}\\
.&\cdots&.\\
.&\cdots&.\\
.&\cdots&.\\
\alpha_{T-nI_2-n,1} & \cdots&  \alpha_{T-nI_2-n,n}
\end{pmatrix}$
\vspace{0.25cm}
\FOR{$i=0$  to  $I_2-1$  step 1 }
\STATE $P \leftarrow$ sort each column of P in descending order
\STATE $s \leftarrow$ $\lfloor \delta\left(I_1+i+1)\right) \rfloor$ 
\STATE $ M[i+1,] \leftarrow$ $P[s,]$
\STATE Add the raw $\left(\alpha_{T-nI_2+i.n,1},\cdots,\alpha_{T-nI_2+i.n,n}\right)$ to $P$
\ENDFOR
\STATE $\text{FIND} \leftarrow$
\vspace{0.5cm}
$\begin{pmatrix}
\alpha_{T-nI_2,1} & \cdots&\alpha_{T-nI_2,n}\\
.&\cdots&.\\
.&\cdots&.\\
.&\cdots&.\\
\alpha_{T-n,1} & \cdots& \alpha_{T-n,n}
\end{pmatrix}$ $\leq$ $M$\\
\STATE \text{FIND$_{i,j}$ takes 1 when $\alpha_{T-n.I_2+i.n,j} \leq M_{i,j}$ and 0 otherwise.}

\vspace{0.5cm}
\hspace*{\algorithmicindent} \textbf{Output:} Percentage of examples in the test set for which the true label is inside the corresponding prediction region : $ \frac{\sum_{i}^{} \sum_{j}^{} FIND_{i,j}}{n.I_2} $; Percentage of examples in the test set for which the $j^{th}$ component of the true label is inside the corresponding prediction interval : $ \frac{\sum_{i}^{} FIND_{i,j}}{I_2} $  ; Mean width : $2.\text{mean}\left(M_{,j}: j=1,\cdots,n\right)$; Median width : $2.\text{median}\left(M_{,j}: j=1,\cdots,n\right)$ 

\end{algorithmic}
\end{algorithm}
To check the validity of our proposed method, we split the time series into training and test datasets. Then, we implement the FPTO-WNN approach to the training dataset to tune the WNN model's parameters: $p^*$ and $k^*$. We apply the proposed method to build the prediction intervals of the points that constitute the test dataset. We check the empirical reliability of the resulting predictive intervals by reporting the percentage of examples in the test dataset such that the true label is inside the corresponding the prediction interval. Finally, we check the efficiency of the method by measuring the tightness of these regions. For this purpose, we use either the mean or the median function. That is, we follow the steps as defined in Algorithm (1) to check the efficiency and the reliability of the proposed method.
Note that, since  $\lfloor \delta\left(I_1+i+1)\right) \rfloor \geq 1$ for $i=\{0,\cdots,I_2-1\}$, $I_1$ must be superior than $\frac{1}{\delta}-1$. In order to build the prediction region of $y_T$, we set $h=I_1+I_2$, $p=p^*$ and $k=k^*$.

\section{Examples }

To evaluate the method derived in Section (\ref{sec.new method}), we investigate the efficiency of the method utilizing two datasets. A simulated time series dataset and a real dataset. We set the size of the test set be equal to 20\% of the available data ($n.I_2=20\%.T$) and the number of the test sets used in FPTO-WNN to be equal to 20\% of training data if $\frac{20\%}{n}.\left(T-nI_2\right) \geq \frac{1}{\delta}-1$ and $\frac{1}{\delta}-1$ otherwise. As mentioned above, $I_1$ must be greater than $\frac{1}{\delta}-1$. Thus, the test set's size must take at least $\frac{1}{\delta}-1$. $I_1$ takes its possible minimal values $\frac{1}{\delta}-1$ and is defined as follows:

\begin{equation}
I_1=\left\{
\begin{array}{l}
  \frac{20\%}{n}.\left(T-nI_2\right), \quad if \quad \frac{20\%}{n}.\left(T-nI_2\right)\geq \frac{1}{\delta}-1 \\
  \frac{1}{\delta}-1, \quad\quad\quad\quad\quad \text{otherwise}
\end{array}
\right.\end{equation}


\subsection{Simulations studies}
In this section we conduct simulations to investigate the performance of our derived approach. We simulate time series data to estimate the theoretical optimal intervals and then compare them with our method's prediction intervals. We simulate monthly data from two models ETS$(A,N,A)$ and ETS$(A,A_d,A)$ where the residual errors follow the standard normal distribution. We set the number of predictions and the confidence level to 3 and 0.95, respectively.
For most ETS models, a prediction interval can be written as : $ \hat{a}_{T+h|T} \pm c \sigma_h$ where $c$ depends on the coverage probability and $\sigma_h^2$ is the forecast variance \citep{hyndman2018forecasting}. Thus, the theoretical widths are calculated as: $2c \sigma_h$. When the forecast errors follow the normal distribution  and using a 95\% confidence level, the constant $c$ is assumed to be equal to 1.96. The forecast variance expressions are :
\[
\begin{split}
    \sigma_h^2 =& \sigma^2\Big[1 + \alpha^2(h-1) + \gamma k(2\alpha+\gamma)\Big]  \hspace{2cm}\text{, for ETS}(A,N,A)\\
    \sigma_h^2 =& \sigma^2\biggl[1 + \alpha^2(h-1) +\gamma k(2\alpha+\gamma)+ \hspace{1.75cm}\text{, for ETS}(A,A_d,A)\\
    &\frac{\beta\phi h}{(1-\phi)^2} \left\{2\alpha(1-\phi) + \beta\phi \right\}-\\
    & \frac{\beta\phi(1-\phi^h)}{(1-\phi)^2(1-\phi^2)} \left\{ 2\alpha(1-\phi^2)+ \beta\phi(1+2\phi-\phi^h)\right\}+\\ &\frac{2\beta\gamma\phi}{(1-\phi)(1-\phi^m)}\left\{k(1-\phi^m) - \phi^m(1-\phi^{mk})\right\}\biggr],
\end{split}
\]
where $\sigma^2$ is the residual variance, m is the seasonal period, $k$ is the integer part of $(h-1)/m$, and $\phi$ is a damping parameter. $\alpha$, $\beta$ and $\gamma$ are the smoothing parameters. For the ETS$(A,N,A)$ model, we generate two series. The first series is of length $T=300$ with the smoothing parameters $\alpha=0.5$ and $\gamma=0.2$. The second series is of length $T=400$ with the smoothing parameters 
$\alpha=0.8$ and $\gamma=0.4$. Similarly, we generate two series from ETS$(A,A_d,A)$. The first series is of length 300 and the model parameters $\alpha=0.7,\beta=0.3,\gamma=0.2, \text{ and } \phi=0.82$. The second series is of length 400 and the model parameters $\alpha=0.8,\beta=0.2,\gamma=0.1, \text{ and } \phi=0.9$ 

From table 1, one can see that most of the empirical widths for the last three predictions are close to the theoretical widths. Moreover, from table 2, one can conclude that the percentage inside predictive regions is high and exceeds the 95 \% confidence level in almost all the scenarios. Overall, the results guarantee the empirical validity and efficiency of our approach.
\begin{table}[H]

\begin{tabular}{|c|c|c|c|c|c|}
  \hline
h & Sample size & \multicolumn{2}{|c|}{ETS$(A,N,A)$}    & \multicolumn{2}{|c|}{ETS$(A,A_d,A)$}  \\
&& Theoretical width& Empirical width & Theoretical width& Empirical width\\
\hline
1&T=300& 3.92& 3.69& 3.92& 3.69\\
& T=400 & 3.92 & 4.01 & 3.92 &4.95 \\
\hline
2&T=300& 4.38 & 4.39& 5.40& 5.34\\
& T=400 & 5.02 &5.73 &5.49 &5.98 \\
\hline
3&T=300& 4.80 & 5.42& 7.02&7.61 \\
& T=400 & 5.92 & 5.90 &7.08 &7.93 \\
\hline
\end{tabular}
\caption{Widths of the prediction intervals for the three last predictions with confidence level 95\% based on our method}\label{tb:table a}
\end{table}

\begin{table}[H]
\begin{tabular}{|c|c|c|c|c|c|}
  \hline
 Model & Sample size & \multicolumn{4}{|c|}{percentage inside predictive regions} \\
&&$ \frac{\sum_{i}^{} FIND_{i,1}}{I_2} $&$ \frac{\sum_{i}^{} FIND_{i,2}}{I_2}$ &$ \frac{\sum_{i}^{} FIND_{i,3}}{I_2} $& $ \frac{\sum_{i}^{} \sum_{j}^{} FIND_{i,j}}{n.I_2} $\\
\hline
ETS$(A,N,A)$ &  T=300 & 100\% & 94.73\% & 100\% & 98.24\%  \\
& T=400 & 100\% & 96.3\% & 96.3\% & 97.53\% \\
\hline
ETS$(A,A_d,A)$ &  T=300 & 94.74\% & 100\% & 100\% &98.25\%  \\
& T=400 & 100\%& 95.45\% & 95.45\% &96.97\%  \\
\hline
\end{tabular}
\caption{Reliability of the proposed method when n=3 and confidence level 95\%}\label{tb:table b}
\end{table}

\subsection{Cow's milk production in the United Kingdom (UK) }

We use the Milk Production data in the UK \citep{eurostat} to evaluate the proposed method derived in Section (\ref{sec.new method}). The time series data  include 634 points ranging from January 1968 to October 2020, and are delineated by month. We implement the proposed algorithm CheckCP in R. To determine the efficiency of conformal prediction based on the WNN approach, we compare it with two underlying algorithms: Autoregressive Integrated Moving Average (ARIMA) and  ETS.   We fit the models on the training set using the functions \textit{auto.arima} and \textit{ets}  from the R package \textit{forecast} \citep{hyndman2018forecast}. Then, we use a similar algorithm \textit{CheckCP}  to measure the validity and the efficiency of conformal predictions using these models. 
\begin{figure}[H]
		\centering
		\mbox{\includegraphics[width=16cm, height=10cm]{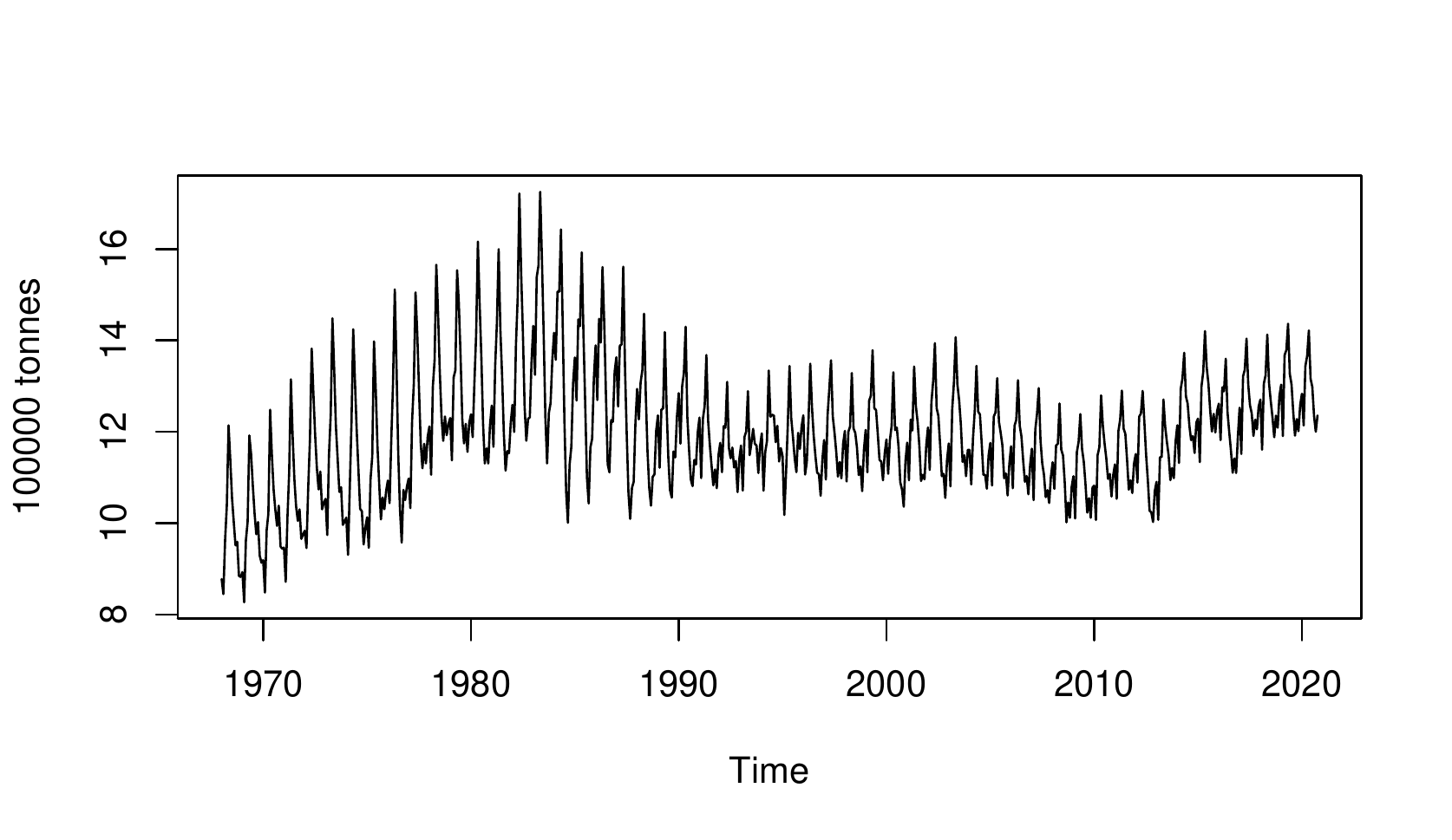}}
		\caption{\label{fig:plot1} Cow's Milk Production in the UK.}
	\end{figure}

Figure 1 depicts the monthly cow’s milk production in a hundred thousand tonnes in the UK and shows the trend across the time. Table 3 presents the MAPE values of the four models on the test dataset for each horizon n \(=1,2,3,4\). All the models provide low MAPE value, which indicates a good forecast accuracy. Overall, our approach outperforms other models across the horizon.
\begin{table}[!ht]
\begin{tabular}{|c|l|l|}
 \hline
 Horizon n & Method & MAPE\\
 \hline
 1 & FPTO-WNN $\left(p^*=11,k^*=7\right)$ & 1.2919 \\
   & SARIMA$\left(2,1,1\right)\left(0,1,1\right)_{12}$ & 1.2269 \\
   & ETS$\left(M,N,A\right) $ & 1.3456 \\
  
    \hline
  2 &  FPTO-WNN$ \left(p^*=6,k^*=6\right)$ & 1.5768 \\
   & SARIMA$\left(2,1,1\right)\left(0,1,1\right)_{12}$ & 1.6621 \\
   & ETS$\left(M,N,A\right) $ & 1.6406 \\
   
   \hline
  3 &  FPTO-WNN$ \left(p^*=4,k^*=7\right)$ & 1.8893 \\
   & SARIMA$\left(2,1,1\right)\left(0,1,1\right)_{12}$ & 1.8608 \\
   & ETS$\left(M,N,A\right) $ & 1.8727 \\
  
   \hline
  4 &  FPTO-WNN$ \left(p^*=2,k^*=10\right)$ & 2.0759 \\
   & SARIMA$\left(2,1,1\right)\left(0,1,1\right)_{12}$ & 2.0703 \\
   & ETS$\left(M,N,A\right) $ & 1.9541 \\
  
   \hline
\end{tabular}
\caption{MAPE values for FPTO-WNN, SARIMA, and ETS methods applied to the Cow's Milk Production data in the UK.}\label{tb:table1}
\end{table}

\begin{table}[H]
\begin{tabular}{|c|l|l|l|l|l|l|}
\hline
 Horizon n & $I_1$ & $I_2$ & Method &\multicolumn{3}{|p{3cm}|}{$\frac{\sum_{i}^{} \sum_{j}^{} FIND_{i,j}}{n.I_2}$} \\
&&&&  $90\%$ &$92\%$& $95\%$\\
 \hline
 1 & 102 & 127 & FPTO-WNN $\left(p^*=11,k^*=7\right)$ & 91.34 &93.70& 96.85 \\
   & & & SARIMA$\left(2,1,1\right)\left(0,1,1\right)_{12}$ & 89.76 & 92.13 & 96.06 \\
   & & & ETS$\left(M,N,A\right) $ & 96.85 & 99.21 & 100 \\
 
    \hline
  2 & 51 & 64 & FPTO-WNN $\left(p^*=6,k^*=6\right)$ & 90.62 &92.19& 95.31\\
   & & & SARIMA$\left(2,1,1\right)\left(0,1,1\right)_{12}$ & 89.06 & 91.41 & 96.87 \\
   & & & ETS$\left(M,N,A\right) $ & 96.09 & 97.66 & 99.22 \\
 
   \hline
  3 & 34 & 43 & FPTO-WNN $\left(p^*=4,k^*=7\right)$ & 92.25 &93.80& 97.67 \\
   & & & SARIMA$\left(2,1,1\right)\left(0,1,1\right)_{12}$ & 87.60 & 91.47 & 95.35 \\
   & & & ETS$\left(M,N,A\right) $ & 95.35& 98.45 & 99.22 \\
  
 \hline
 4 & 26 & 32 & FPTO-WNN $\left(p^*=2,k^*=10\right)$ & 91.40 &94.53& 96.90 \\
 & & & SARIMA$\left(2,1,1\right)\left(0,1,1\right)_{12}$ & 86.72 & 91.41 & 93.75 \\
 & & & ETS$\left(M,N,A\right) $ & 94.53 & 96.87 & 98.44 \\

 \hline
\end{tabular}
\caption{Reliability of the obtained prediction regions for FPTO-WNN, SARIMA, and ETS  models using Cow’s Milk Production dataset in the UK.}\label{tb:table2}
\end{table}

Table 4 presents the reliability of the obtained prediction regions for each underlying algorithm. $\frac{\sum_{i}^{} \sum_{j}^{} FIND_{i,j}}{n.I_2}$ finds the percentage of the true label inside the predictive regions. It reports the percentage of examples for which the true label is inside the region output by each method. It is clear that for the confidence levels 90\%, 92\% and 95\%, the underlying algorithms are empirically valid since the percentages reported are close to and exceed, in some cases, the confidence levels. Tables 5, 6, 7, and 8 report the reliability and the tightness of the three methods using 1, 2, 3, and 4 components, respectively. It is clear that the models with high percentage inside predictive regions , such as the ETS, have low efficiency. Also, one can see that, in general, the FPTO-WNN method provides a low mean width and a low median width. That is,  the FPTO-WNN method is the most efficient method.

\begin{table}[H]
\begin{tabular}{|c|c|c|c|c|c|c|c|c|c|}
  \hline
Method & \multicolumn{3}{|p{3cm}|}{Mean width}    & \multicolumn{3}{|p{3cm}|}{Median width } & \multicolumn{3}{|p{3cm}|}{ $\frac{\sum_{i}^{} FIND_{i,j}}{I_2} $} \\
&  $90\%$ &$92\%$& $95\%$ & $90\%$ &$92\%$& $95\%$ & $90\%$ &$92\%$& $95\%$\\
\hline
FPTO-WNN & 0.633 & 0.692 &  0.823 & 0.634 &0.692& 0.836 &91.34&93.70& 96.85  \\
SARIMA &0.625 & 0.681 & 0.840& 0.626& 0.686&0.840&89.76 & 92.13&96.06   \\
ETS  & 0.889 &0.982 & 1.141 & 0.886 & 0.957 & 1.130 & 96.85 &99.21 & 100   \\

    \hline
\end{tabular}
\caption{Reliability and the tightness of FPTO-WNN, SARIMA, and ETS methods when $n=1$ using the Cow’s Milk Production data in the UK dataset.}{\label{tb:table3}}
\end{table}

\begin{table}[!ht]
\begin{tabular}{|c|l|l|c|c|c|c|c|c|c|c|}
  \hline
 $j^{th}$ & Method & \multicolumn{3}{|p{3cm}|}{Mean width}    & \multicolumn{3}{|p{3cm}|}{Median width } & \multicolumn{3}{|p{3cm}|}{ $\frac{\sum_{i}^{} FIND_{i,j}}{I_2} $} \\
  
& &  $90\%$ &$92\%$& $95\%$ & $90\%$ &$92\%$& $95\%$ & $90\%$ &$92\%$& $95\%$\\
\hline
1 & FPTO-WNN & 0.692 & 0.756 &  0.881 & 0.726 &0.769 & 0.855 &87.50 &90.62 &93.75 \\
& SARIMA & 0.730 & 0.775  & 0.979& 0.744& 0.787 & 1.016 & 87.50 &89.06 &96.87 \\
& ETS  & 0.990 & 1.096 & 1.248 &0.948&1.103  & 1.261 & 98.44 & 98.44& 100   \\

\hline
2 & FPTO-WNN &0.822  & 0.990 &  1.244 & 0.827 &1.066 & 1.253 &93.75 &93.75 & 96.87  \\
& SARIMA & 1.064 & 1.164  & 1.401 & 1.092& 1.162 & 1.375 &90.62  & 93.75 & 96.87  \\
& ETS  & 1.125 &  1.155&  1.268 & 1.119 & 1.143 & 1.207 &93.75 & 96.87&  98.44  \\

 \hline
\end{tabular}
\caption{Reliability and the tightness of FPTO-WNN, SARIMA, and ETS methods when $n=2$ using the Cow’s Milk Production data in the UK dataset.}{\label{tb:table4}}
\end{table}
\vspace{2cm}

\begin{table}[H]
\begin{tabular}{|c|l|l|c|c|c|c|c|c|c|c|}
  \hline
  	 $j^{th}$ & Method & \multicolumn{3}{|p{3cm}|}{Mean width}    & \multicolumn{3}{|p{3cm}|}{Median width } & \multicolumn{3}{|p{3cm}|}{  $ \frac{\sum_{i}^{} FIND_{i,j}}{I_2} $} \\
  
  	   & &  $90\%$ &$92\%$& $95\%$ & $90\%$ &$92\%$& $95\%$ & $90\%$ &$92\%$& $95\%$\\
  	  \hline
  	  
  	 1 & FPTO-WNN & 0.676 & 0.737 &  0.825 & 0.679 & 0.682& 0.811 & 90.70 & 90.70 & 95.35\\
   & SARIMA & 0.625 & 0.709 & 0.837 & 0.636& 0.744 & 0.822 & 88.37 & 93.02& 95.35\\
 & ETS  & 0.840 & 0.960 & 1.127 &0.848 &  0.920 & 1.130 & 93.02 & 100 & 100   \\
  
    \hline
  2 & FPTO-WNN & 0.862  & 1.039 &  1.291 & 0.844 & 1.127 &  1.289& 93.02& 95.35 &100   \\
   & SARIMA & 1.090 & 1.171  & 1.215 &1.125 & 1.162 &1.210  & 93.02  & 93.02 & 97.67  \\
 & ETS  & 1.194 & 1.303 & 1.523  & 1.126 & 1.337 & 1.541 & 97.67& 100&100    \\
  
 \hline
 3 & FPTO-WNN & 1.184 &1.262 & 1.441  & 1.198 & 1.259& 1.361 &93.02 &95.35  &97.67 \\
   & SARIMA & 1.165 &  1.303 & 1.526 &1.196 & 1.394 & 1.565 & 81.39 &88.37 & 93.02\\
 & ETS  & 1.377 & 1.428 & 1.605 &1.407 &  1.407 & 1.584 &95.34 & 95.35 &  97.67  \\
  
    \hline
\end{tabular}
\caption{Reliability and the tightness of FPTO-WNN, SARIMA, and ETS methods when $n=3$ using the Cow’s Milk Production data in the UK dataset.}{\label{tb:table5}}
\end{table}

\begin{table}[H]
\begin{tabular}{|c|l|l|c|c|c|c|c|c|c|c|}
  \hline
 $j^{th}$ & Method & \multicolumn{3}{|p{3cm}|}{Mean width}    & \multicolumn{3}{|p{3cm}|}{Median width } & \multicolumn{3}{|p{3cm}|}{ $ \frac{\sum_{i}^{} FIND_{i,j}}{I_2} $} \\
 & &  $90\%$ &$92\%$& $95\%$ & $90\%$ &$92\%$& $95\%$ & $90\%$ &$92\%$& $95\%$\\
 \hline
 1 & FPTO-WNN & 0.777 & 0.842 &  0.935 & 0.765 & 0.889& 0.924 & 93.75 & 93.75 & 96.87\\
 & SARIMA &0.848  & 0.997  & 1.136 &0.787&  1.016& 1.1056 & 90.62 &100 &100 \\
 & ETS  & 0.912 & 1.059 & 1.387 &0.948 &  0.957 & 1.261 & 96.87 &100  &100    \\
 
 \hline
  2 & FPTO-WNN & 1.062  & 1.135 &  1.488 & 1.027 & 1.132 & 1.353 &90.62 & 96.87 & 96.87  \\
 & SARIMA & 1.240 &  1.435 & 1.596 & 1.180& 1.418& 1.506 & 93.75  &  93.75& 100  \\
 & ETS  & 1.131 & 1.156 &  1.326 & 1.119 & 1.143 & 1.187 &93.75 &100 &100    \\
  
 \hline
 3 & FPTO-WNN & 1.182 & 1.291 &  1.555 & 1.143 &1.195& 1.591 & 93.75& 96.87 &96.87 \\
 & SARIMA & 1.096 & 1.169 & 1.320 & 1.123& 1.198 & 1.413 & 84.37 &84.37 &84.37 \\
 & ETS  & 1.422 & 1.513 & 1.592 &1.363 &  1.465 & 1.595 & 96.87 & 96.87 & 96.87   \\
 
    \hline
  4 & FPTO-WNN & 1.551 & 1.687 & 1.883  & 1.675 & 1.709&1.922  & 87.50 & 90.62 &96.87 \\
   & SARIMA & 1.281 &  1.383 & 1.543 &1.369 & 1.417 & 1.631 & 78.12 &87.50 &90.62 \\
 & ETS  & 1.547 & 1.643 & 1.844 &1.519 &  1.601 & 1.853 & 90.62 & 90.62 &  96.87  \\
  
    \hline
\end{tabular}
\caption{Reliability and the tightness of FPTO-WNN, SARIMA, and ETS methods when $n=4$ using the Cow’s Milk Production data in the UK dataset.}{\label{tb:table6}}
\end{table}

\newpage
\section{Conclusion}

This paper addressed the problem of forecasting prediction intervals. We used the conformal prediction and applied the weighted nearest neighbors based on the fast parameters tuning technique in the weighted nearest neighbors as its underlying algorithm. In particular, we showed how to extend the conformal prediction in time series data and how to apply it with nearest neighbors in the multidimensional case. We introduced a new algorithm that checks the empirical validity and efficiency of the proposed method. With the violation of the exchangeability assumption, the results on the simulated time series data and the Cow's Milk Production dataset in the UK showed that the conformal prediction can still end up to be empirically valid and efficient. Moreover,  implementing the WNN method as the underlying algorithm leads to a tight regions  with efficient results. \\ \\

\noindent \textbf{Data Availability Statement} :\\
For our analysis, Cow's Milk Production in the UK are provided by Eurostat, which are available at:\\  http://appsso.eurostat.ec.europa.eu/nui/show.do?dataset=apro\_mk\_colm\&lang=en

\bibliographystyle{asa}
\bibliography{references}
 \end{document}